\documentclass[a4paper]{spie}  

\usepackage{amsmath,amsfonts,amssymb}
\usepackage{graphicx}
\usepackage[colorlinks=true, allcolors=blue]{hyperref}

\title{Charge migration characterization in the METIS H2RG detectors}

\author[a]{Danny Gasman}
\author[b]{Benoît Serra}
\author[a]{Roy van Boekel}
\author[c]{Ioannis Argyriou}
\author[d]{Daniel Dicken}
\author[a]{Vianak Naranjo}
\author[a]{Peter Bizenberger}
\affil[a]{Max-Planck-Institut für Astronomie, Königstuhl 17, 69117 Heidelberg, Germany}
\affil[b]{European Southern Observatory, Address, City, Germany}
\affil[c]{Institute of Astronomy, KU Leuven, Celestijnenlaan 200D, B-3001 Leuven, Belgium}
\affil[d]{UK Astronomy Technology Centre, Edinburgh EH9 3HJ, UK}

\authorinfo{Further author information: (Send correspondence to D.G.)\\D.G.: E-mail: dagasman@mpia.de}

\begin{document}
\maketitle

\begin{abstract}
The Mid-infrared ELT Imager and Spectrograph (METIS) is one of the first-light instruments of the Extremely Large Telescope (ELT). For the L and M band, the instrument makes use of Teledyne’s H2RG detectors, known to be affected by charge migration, also known as the brighter-fatter effect (BFE). In the H2RG detectors this manifests as photo-electrons moving from a central bright pixel to neighboring pixels as the depletion region of the central pixel shrinks with accumulated charge. Due to its effect of ‘blurring’ information, it especially affects direct imaging, transit, and high-resolution spectroscopy applications. We aim to characterize its effect in the METIS H2RG detectors, such that we can remove it without losing information.
\end{abstract}


\section{INTRODUCTION}
\label{sec:intro}  

The Mid-infrared ELT Imager and Spectrograph (METIS) is one of the first-light instruments for the ESO Extremely Large Telescope (ELT). It provides a variety of scientific modes in the mid-infrared: an L, M, and N band imager with a long-slit spectropgraph (LSS), and a high-resolution integral field spectrometer in L and M band (LMS), all supported by an internal single conjugate adaptive optics system (SCAO). Both the Imager and LMS include high-contrast imaging (HCI) modes, enabled by coronagraphs.

To allow METIS to be sensitive to all three bands, the instrument makes use of two different detector types. The L and M band are covered by Teledyne's H2RG HgCdTe detectors with a 5.3~$\mu$m cut-off, whereas the N band is covered by a HgCdTe GeoSnap detector with a 13~$\mu$m cut-off. The H2RGs have significant heritage in astronomical applications, being part of numerous ground- and space-based telescopes. A total of five H2RGs are included in the METIS instrument; one in the Imager, and a mosaic of four in the LMS.

In this work, we focus on the characterization of the science-grade H2RG detector with the largest well-depth, most suited to be used in the Imager. The H2RGs have 2048x2048 pixels with a size of 18~$\mu$m, mapped on-sky to 5.47~mas per pixel in the Imager. The detectors contain a donor (n-type) and acceptor (p-type) side, induced by a positively and negatively charged HgCdTe layer, resulting in a voltage difference.\cite{ref:07Ri} When a photon hits the donor lattice, it may free up a charge carrier in the form of a hole-electron pair. They can diffuse through the material, until the hole reaches the depletion region caused by the contact voltage in the detector. The depletion region carries the electron across to the acceptor side. The lattice is then connected to the read-out integrated circuit (ROIC) through indium bumps. The ROIC amplifies the signal and a voltage change is measured. The analog-to-digital converter (ADC) converts the signal to an Analog-to-Digital Unit (ADU). Four rows/columns of reference pixels on all sides of the 2048x2048 pixel grid can be used to measure drift in the amplifiers.

There are various effects causing the detectors to exhibit non-linear behavior during the accumulation of signal. Here, we focus on what we will refer to as `classical non-linearity,' an effect that causes the number of electrons recorded to decrease as charge is accumulated. The accumulation of charge results in a decrease in depletion region size, meaning not all hole-electron pairs will reach the p-n junction, and will not be detected.\cite{ref:07Ri} The rate of incoming photons does not change, but the signal registered by the detector does. This shrinking of the depletion region has a secondary effect, which is of importance to the work presented here. In case of varying incoming photon flux across the detector, neighboring pixels may see a difference in depletion width. This causes an effective change in pixel boundaries, where a more brightly illuminated pixel appears `smaller', resulting in a higher probability of a hole-electron pair to diffuse towards and be captured by the fainter neighboring pixel.\cite{ref:17PlShSm} This process of migrating charge is often referred to as the brighter-fatter effect (BFE), due to its broadening effect on the instrumental point-spread function with accumulating charge. It should be noted, however, that it is (at least to first order) not a function of flux, but rather a function of charge accumulation relative to pixel well-depth. This means that charge migration is \textit{also} important for faint sources, provided the exposure is long enough.

While now being a known problem in multiple detector types, the effect of migrating charge broadening PSFs was first observed in CCDs,\cite{ref:06DoBaSi} where a local electric field in a saturating pixel can repel newly generated electrons to neighboring pixels.\cite{ref:15GuAsAn} This repelling effect can result in multi-pixel migration distances. The migration in the H2RG detectors, being primarily a diffusion effect, is largely limited to direct neighbors \cite{ref:17PlShSm}. This manifestation of charge migration is similar to the Si:As impurity band conduction (IBC) detectors historically used in instruments operating in longer infrared wavelengths than METIS, such as the Mid-InfraRed Instrument (MIRI) on the James Webb Space Telescope (JWST). In this case, the detector non-linearity is also strongly coupled to the charge migration effect, but limiting diffusion distances prior to recombination result in attraction by a neighboring pixel's electric field to become the driving mechanism \cite{ref:23ArLaRi}.

Despite its notoriety, charge migration is not always corrected for in instrument pipelines. It requires adequate knowledge of the physics, and more specifically, of the magnitude of the effect in individual units of the same detector type. Each detector is unique, meaning the numbers required for corrections are slightly different. Especially the performance of METIS, with its high spectral resolution mode and high-contrast imaging, can become significantly limited by migrating charge. Furthermore, basic characterization of detector effects often relies on flat fields, something that becomes impossible to do for the LMS once all detectors have been integrated into the final instrument. The LMS employs image slicers to split and subsequently disperse different parts of the sky, meaning the observations will never be flat once the detectors are installed in the final configuration. It is therefore imperative that we do as much as possible during the detector lab tests at ESO.

\begin{figure}[h]
    \centering
    \includegraphics[width=0.9\linewidth]{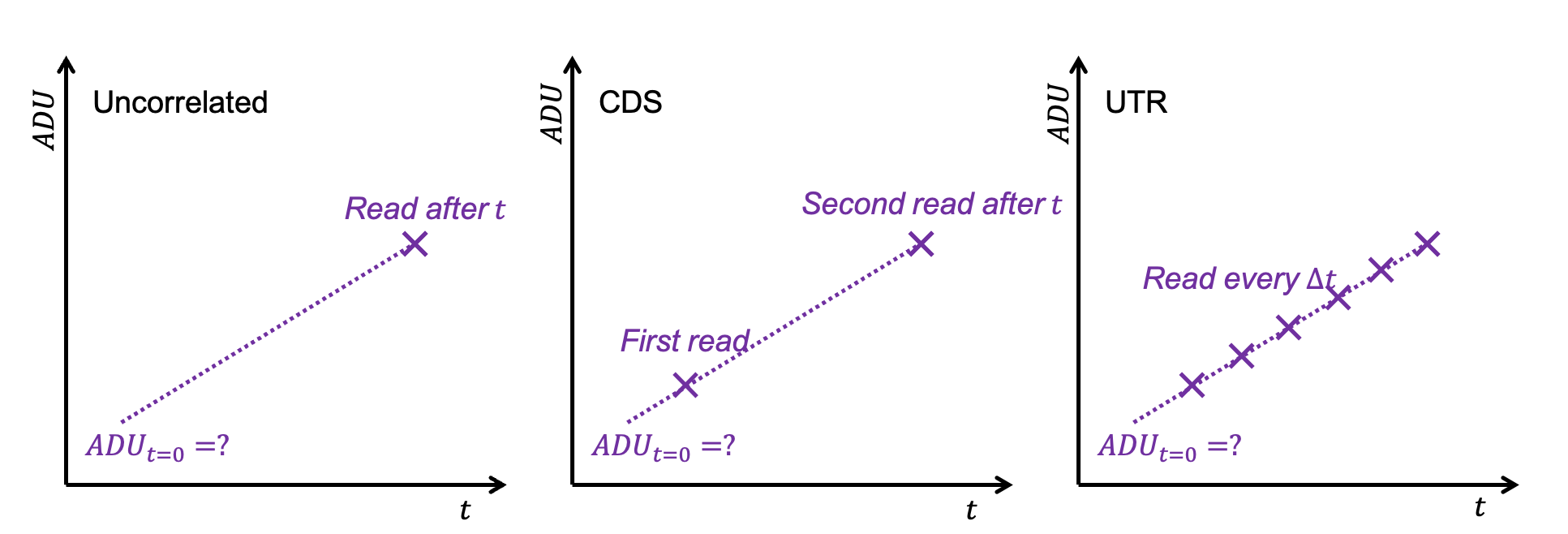}
    \caption{Schematic of the possible sampling schemes. Uncorrelated sampling is illustrated on the left, CDS in the centre, and UTR on the right.}
    \label{fig:sampling}
\end{figure}

The advantage of the HxRG detector series in this context, is that the detectors can be read out non-destructively. The different modes of METIS may operate using different read-out schemes. Extremely bright objects observed with the Imager may benefit from the uncorrelated read-out, illustrated on the left in Fig.~\ref{fig:sampling}, where a single frame is read after a specified amount of time prior to resetting the pixels. While this allows for the shortest integration times (or `DITs'), the disadvantage of this mode is that the precise pixel bias is not known. A next option is then to get some measure of this bias by sampling twice during a cumulative exposure, shown in the centre of Fig.~\ref{fig:sampling}, once at the start and once at the end, prior to resetting. This is referred to as correlated double sampling (CDS), and is currently set to be the standard for most of the scientific observing modes of METIS. While the smaller number of samples result in a lower data volume, any undesirable behavior of the detector during the accumulation of signal cannot be traced. This means that pixels that saturate at some unspecified time during the cumulative measurement cannot be recovered. The final option, then, is to instead regularly measure samples during the non-destructive accumulation of signal, shown on the right in Fig.~\ref{fig:sampling}. This is referred to as up-the-ramp (UTR) sampling. We note that the detectors are to be controlled by the NGCII controller of ESO, which can automatically subtract reference pixels and fit ramps prior to writing the data. When turning this option off, the read-out scheme is referred to as UTR-raw.

For the purpose of the detector characterization presented in this work, we exclusively make use of UTR-raw sampling, but will refer to this simply as UTR henceforth. The advantage of UTR is that any temporal variations in the detector behavior can be traced directly. Therefore, unlike in CCDs, one does not need to measure a series of flats of increasing exposure time to sample the non-linear behavior of the detector.

We structure this work as follows. The lab test set up at ESO is described in Sect.~\ref{sec:test}. We summarize the theory behind the detector characterization based on fundamental work done for the \textit{Roman Space Telescope} in Sect.~\ref{sec:char},\cite{ref:HiChoI,ref:ChoHiII,ref:20FrGiCh} and the results of these measurements in Sect.~\ref{sec:results}. Finally, we discuss future testing and characterization work in this context in Sect.~\ref{sec:summary}.

\section{LAB TESTS AT ESO}
\label{sec:test}

The individual METIS detectors are tested for compliance with their required performance in the Mosaic Test Facility (MTF) at ESO Headquarters in Garching, Germany. The test setup is kept constant between detectors, which are tested at representative conditions by being cryogenically cooled to 40~K.

For the purpose of the characterization activities presented here, a set of 40 dark exposures and a set of 40 flat exposures are taken. Both of these sets are 150~s exposures with 102 UTR frames. For the flats, the detectors are illuminated using a 380~K cryogenic blackbody. The detectors are controlled using ESO's new NGCII controller. The detectors can be operated in slow (100~kpix/s) or fast (5~Mpix/s) read-out mode. Slow mode was used for the test data presented here. Furthermore, the detectors can be reset by row (row reset, RR), or globally (global reset, GR). In these tests, the detectors are reset row-by-row.

\begin{figure}[h]
    \centering
    \includegraphics[width=0.9\linewidth]{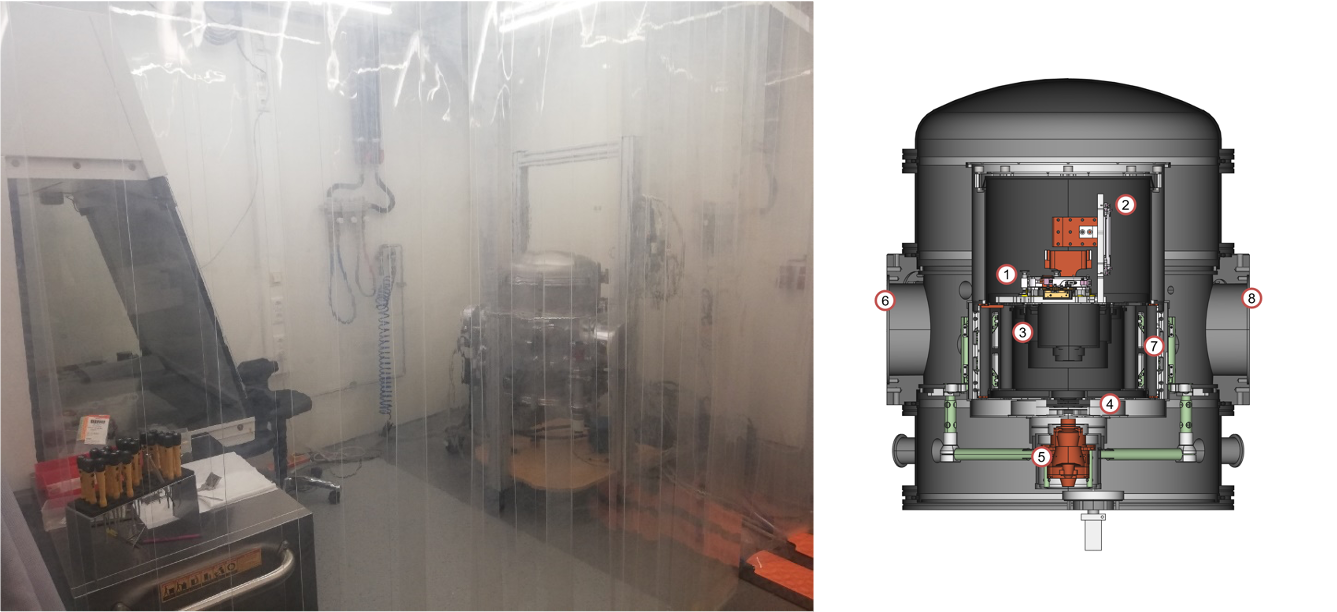}
    \caption{Picture of the test lab at ESO (left), and a schematic of the cryogenic detector mount (right).}
    \label{fig:eso_test}
\end{figure}

Finally, the voltages applied to the detectors can change a variety of the detector properties. We note that the data presented here assume the values recommended by the detector manufacturer Teledyne Imaging Systems, but these will be tuned in the near future to reach the desired behavior. Future tests will therefore not be done at the same settings to repeat and expand on the analysis presented here. This means that exact results of the characterization presented here must be assumed to be preliminary. The applied methods and lessons learnt, however, will be important for future tests.

\section{THEORY OF CHARACTERIZATION}
\label{sec:char}

The quantification of charge migration in the H2RG detectors has been done in various ways in the recent past. One is the use of flat correlations,\cite{ref:HiChoI} which is similar to methods applied to CCDs,\cite{ref:14AnAsDo} whereas another method is the use of spot measurements \cite{ref:24PlShCh}. It should be noted that an 18\% discrepancy was found between the results of both methods, although the exact cause has not been investigated in detail.\cite{ref:24PlShCh} Given the setup available at ESO, we opted for the use of flat fields, as projecting an unresolved point source on the detector is currently not possible. As mentioned above, a set of 40 darks and flats was taken. Before we can use any of these data sets in an autocorrelation measurement, some preprocessing steps must be applied. We outline these first, before explaining the characterization of charge migration.

\subsection{Reference pixels}
As mentioned above, the drift in the amplifiers can be measured through a series of unilluminated pixels along the edges of the detectors. In nominal operations, the NGCII controller will automatically use these reference pixels to correct for this drift in the data. Since this was not applied in our test data, we had to do this manually prior to further processing. It was previously found that row-based reference pixel subtraction introduced extra noise into the data. Therefore, we exclusively used an average of the reference pixels column-by-column for the correction. Due to a glow at the top of the detector (further discussed in Sect.~\ref{sec:darks}), we omitted these rows of reference pixels from the average. Tuning of the sampling speed may allow us to use these pixels in the future.

\subsection{Inter-pixel capacitance}
\label{sec:ipc_method}
After applying the reference pixel correction on each data set, the next thing to address was inter-pixel capacitance (IPC). IPC is a form of cross-talk that may be confused with charge migration, as it also causes pixels to be correlated with each other. However, IPC is primarily a readout effect, namely parasitic capacitance between different readout channels. While we assume this to be constant in this work, a non-linear IPC component has been proposed to affect H2RG data.\cite{ref:16DoNiBa}

A way to measure IPC in detectors that does not rely on autocorrelations, is by using isolated hot pixels in masterdarks.\cite{ref:11HiMc} We combined the set of reference pixel-corrected darks to create this masterdark and improve the S/N on this measurement. A background signal was estimated in the hot pixel area, excluding a 5x5 pixel square around the hot pixels. After subtracting this signal, we could find the signal registered in the neighboring pixels compared to the hot central pixel. This is converted to a percent measurement by normalizing the 5x5 square by its sum. Under the assumption that the hot pixel is not registering any `real' signal, which we can expect to be the case at 40~K, any signal registered in the neighbors is due to cross-talk in the electronics.

Once the IPC was characterized, we could use the results to create a deconvolution kernel and deconvolve our data sets. We used four iterations of Richardson-Lucy deconvolution, which is typically used for PSF deconvolution, to correct the IPC effect.\cite{ref:72Ri,ref:74Lu}

\subsection{Dark subtraction}
The masterdark that was used for the IPC characterization, was also used for dark subtraction from the flat data prior to non-linearity fitting. The average of the darks was calculated to retrieve a high S/N dark measurement. This removed the influence of the dark current on the data.

\subsection{Non-linearity}
\label{sec:linearity_method}
Next, we had to characterize the classical non-linearity behavior of the detector. This is done using the flats, under the assumption that charge migration evens out in flat illumination, and the classical non-linearity measurement is not significantly affected. A common approach to measure this, is by fitting a set of polynomials. The first is to the ramps themselves, leaving ample margin prior to saturation:
\begin{equation}
    ADU(t) = C_0 + C_1 t + C_2 t^2 \text{.}
    \label{eq:ramp_fit}
\end{equation}
The assumed perfectly linear ramp was then
\begin{equation}
    ADU_{lin}(t) = C_0 + C_1 t \text{.}
    \label{eq:lin_fit}
\end{equation}
The deviation from linearity is then described as a function of accumulated $ADU$, under the assumption that it is solely a function of well-depth. This final correction curve then becomes
\begin{equation}
    ADU_{lin}(t)/ADU_{raw}(t) = C_{l0} + C_{l1} ADU_{raw}(t) + C_{l2} ADU^2_{raw}(t) \text{.}
    \label{eq:nonlin_fit}
\end{equation}
Since the first term implies the ratio between the ideal and observed ramps when no charge has been collected yet, it should be equal to 1 in an ideal case. We therefore normalized the other coefficients in the fit results by $C_{l0}$. Due to differences in the lattice and other properties of individual pixels, these coefficients can be different across the detector. There is a trade-off between accurate non-linearity corrections per pixel, and high S/N on the measurement. We therefore characterize this per `super-pixel', similarly to our charge migration measurement, as we will outline below.

\subsection{Correlation function}
For this part of the work, we follow pre-defined methods based on examining the correlations between pixels going up the ramp and with their neighbors in flats.\cite{ref:HiChoI,ref:ChoHiII} This method makes use of the fact that UTR sampling allows for temporal as well as spatial information in a single exposure. The idea is that, in case of charge migration, the samples depend on values of previous samples, as well as those in the neighbors. The correlation function of interest to solve is
\begin{equation}
    [K^2a' + KK']_{\Delta i,\Delta j} = \frac{g^2}{I^2 t_{ab} t_{cd}} C_{abcd}(\Delta i,\Delta j) + \left\{ 
    \begin{aligned}
        2(1-8\alpha)\beta \hspace{0.5cm} (\Delta i,\Delta j)=(0,0) \\
        4 \alpha_{H} \beta \hspace{0.5cm} (\Delta i,\Delta j)=(\pm 1,0) \\
        4 \alpha_{V} \beta \hspace{0.5cm} (\Delta i,\Delta j)=(0,\pm 1) \\
        0 \hspace{0.5cm} \text{for all other $(\Delta i, \Delta j).$}
    \end{aligned}
\right.
\label{eq:kfunc}
\end{equation}
Here, kernel $K$ is the linear IPC kernel, whereas kernel $K'$ incorporates non-linear, signal dependent, IPC contributions (NL-IPC). While we assumed that the IPC is linear in our hot pixel analysis (an assumption that we will discuss the validity of later), we find the total combined kernel $[K^2a' + KK']_{\Delta i,\Delta j}$ from the above equation, and stopped our analysis there in this work. The charge migration related term is $a'$, which describes the kernel coefficients. The $(\Delta i,\Delta j)$ indicates $x$- and $y$-pixel coordinates relative to a central pixel. For charge migration in H2RG detectors, it is typically assumed that its effect are limited to $\pm$2 pixels around a central pixel. $g$ and $I$ indicate the gain and current in units $e^-/ADU$ and $e^-/s$, respectively. This means that their combined quantity reduces to the rate in $ADU/s$, which we can use instead. $t$ is the elapsed time between frames, where $a$, $b$, $c$, and $d$ indicate different frame numbers. In case of the non-overlapping correlation examined here, we assume that $a < b < c < d$. We also take equal intervals up the ramp, meaning $b-a = d-c$. Finally, the last term is related to the combined non-linearity and IPC contribution. We have already quantified this using the method from Sect.~\ref{sec:linearity_method}. $\alpha$ is the mean of all neighboring components of the IPC kernel, whereas subscripts $V$ and $H$ denote the mean of the vertical and horizontal neighbors, respectively, which we based on the hot pixel analysis discussed above. Finally, $\beta$ is the first-order non-linearity term, equivalent to\cite{ref:17PlShSm}
\begin{equation}
    \beta = \frac{C_2}{C_1^2}
\end{equation}
from Eq.~\ref{eq:ramp_fit}.

This then leaves the correlation function $C$. We had a series of flats available to us, which has the added benefit that we can difference them to remove pixel-specific variations, such as small variations in pixel size. There are many permutations in which to do the differencing. We opted for doing this in series with time, in other words
\begin{equation}
    \Delta S_{n+1,n} = S_{n+1} - S_{n} \text{.}
\end{equation}
The correlation function between all these flats then becomes
\begin{align}
\begin{aligned}
    \Delta C_{abcd}(\Delta i,\Delta j) = \frac{1}{N_{pair}} \sum_{n}^{N_{pair}} \frac{1}{XY} \sum_{i,j}^{X,Y} \left \{ \left( [S_{n_b} - S_{n_a}](i,j) - \overline{[S_{n_b} - S_{n_a}]} \right) \right. \\ \left. \times \left([S_{n_d} - S_{n_c}](i+ \Delta i,j+ \Delta j) - \overline{[S_{n_d} - S_{n_c}]} \right) \right \} \text{,}
\end{aligned}
\label{eq:corrfunc}
\end{align}
where $N_{pair}$ is the number of flat pairs (in our case $40/2=20$), and $X$ and $Y$ denote the total number of pixels in the $x$ and $y$-directions of the detector, such that the last sum in the expression is equivalent to an expected value $E$.

Finally, the analysis can be done in super-pixels, to increase the S/N on the measurement. A smaller super-pixel allows for better tracing of spatial variations across the detector, but has a lower S/N than a larger super-pixel. Excluding reference pixels, the H2RG has 2040x2040 pixels. We divide this into 17x17 super-pixels of 120x120 pixels.

\section{CHARACTERIZATION RESULTS}
\label{sec:results}
We applied the methods described above to the METIS LM-Imager science detector. The flats and darks are from an initial check, meaning the exact voltages under which we will operate the detector in the end are subject to change. The numbers and behavior cited in this section are therefore not final. Regardless, it provides a first step towards characterizing and eventually correcting this effect. We step through the results in the same order as Sect.~\ref{sec:char}.

\subsection{Darks and reference pixels}
\label{sec:darks}

\begin{figure}[h]
    \centering
    \includegraphics[width=0.9\linewidth]{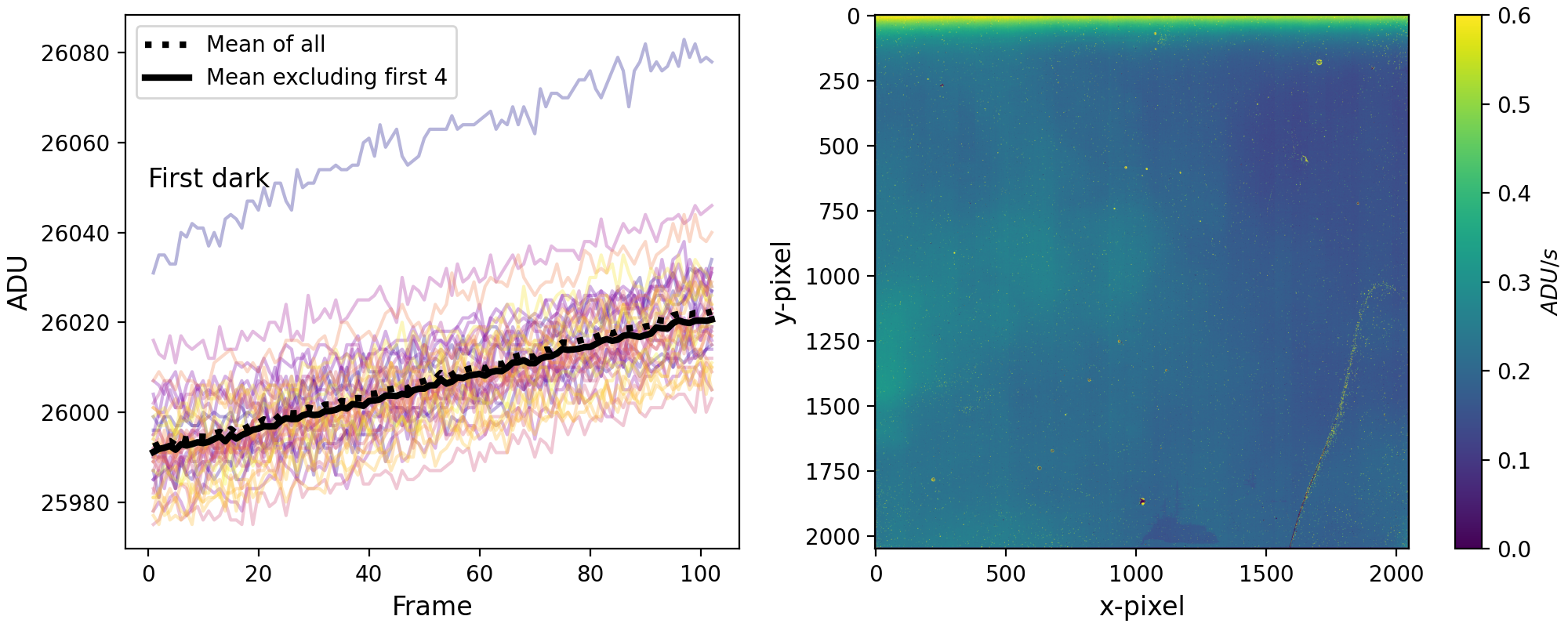}
    \caption{Selected pixel ramps from all 40 darks (left), and the resulting masterdark (right). The first exposure was significantly brighter than the others. The black dotted line is the mean when including all darks, whereas the solid line is the mean when ignoring the first four. The masterdark on the right is the result of the solid line.}
    \label{fig:dark_anomaly}
\end{figure}

We observed a transient in the darks, where the first was clearly brighter than the rest, and attribute this to settling effects. This offset is shown in the left panel of Fig.~\ref{fig:dark_anomaly}, which also demonstrates the offset of the masterdark when including all 40 ramps. While this difference is small, to ensure the darks are properly dark, we omit the first four darks from the masterdark. The resulting masterdark is shown on the right in Fig.~\ref{fig:dark_anomaly}. Some areas of bad pixels can be seen, along with a characteristic glow at the top of the detector.

\begin{figure}[h]
    \centering
    \includegraphics[width=0.9\linewidth]{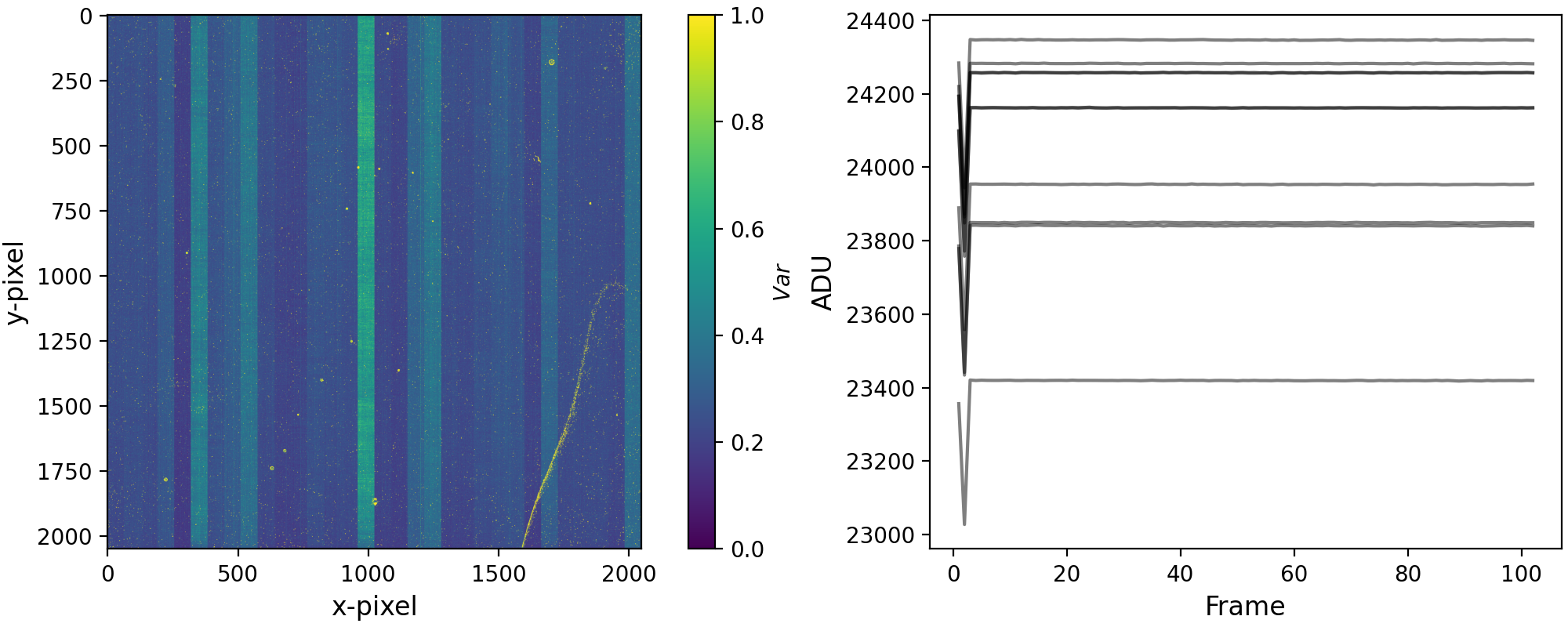}
    \caption{Variance of the ramps in the masterdark (left), and a sample of reset pixel ramps from the top of the detector (right). We note that the variances were calculated using frame numbers 3 and onwards, to omit the influence from the spike at the start of the ramp. This feature is not present in the reference pixels at the bottom of the detector. Clusters of bad pixels also clearly show up, with the most notable one the hooked-feature on the bottom right of the image.}
    \label{fig:variance_glow}
\end{figure}

The glow in HxRGs can be caused by thermal behavior of the detector, or by the ADC. In case of the latter, these pixels should have no excess variance compared to the rest of the detector. In fact, for H2RGs, the glow is not expected to be thermal in origin. This is demonstrated by the lack of excess variance observed in the ramps of the reference pixels at the top of the detector, shown on the left in Fig.~\ref{fig:variance_glow}. The observed increase in signal on the right of Fig.~\ref{fig:dark_anomaly}, is not actually an increase in signal, but rather an artefact of an anomaly at the start of the reference pixel ramps, visible in the right panel of Fig~\ref{fig:variance_glow}. Due to this, a slightly higher rate will be derived from a linear fit or mean of the frame differences. However, when calculating the variance on these ramps starting after this negative spike, it is similar to those at the bottom of the detector. Regardless, due to this observed effect, we only use the reference pixels at the bottom for our reference pixel correction.

In addition to the behavior of the reference pixels, the variance map also shows vertical striping, corresponding to the 32 read-out channels of the detector. It is apparent that some of these introduce more noise into the data than others. We keep this in mind as we define our limits for bad pixel detection, which we do on the masterdark. This is especially important for the IPC characterization.

\subsection{IPC}
\begin{figure}[h]
    \centering
    \includegraphics[width=0.6\linewidth]{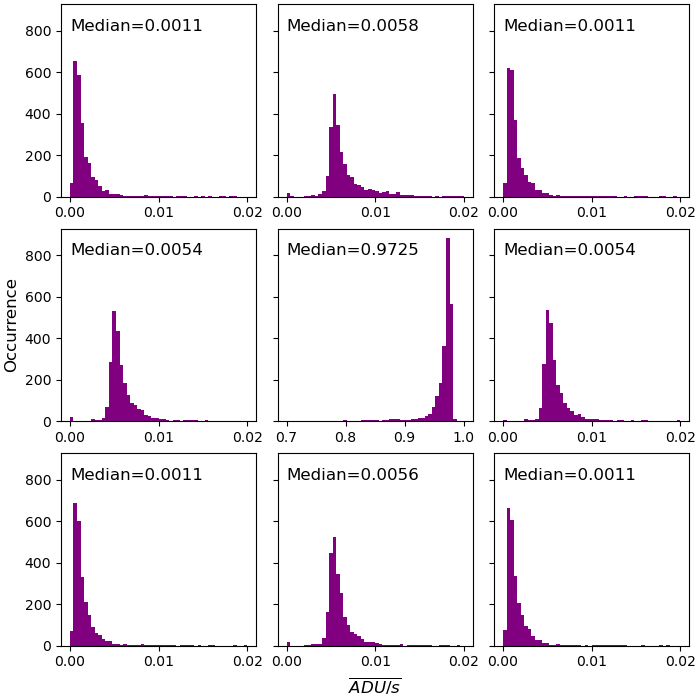}
    \caption{Statistics of the linear IPC analysis based on isolated hot pixels. The median values of the distributions used in the IPC correction kernel are shown per panel.}
    \label{fig:ipc_kernel}
\end{figure}

In order to characterize the IPC, we need to find hot pixels in the masterdark. We defined these as pixels that had 5$\sigma$ more signal than the median. Once these pixels were identified, we ensured they were isolated by checking six pixels on all sides for hot pixels. If there was one, or more, these pixels were excluded from the analysis. We identified 5468 isolated hot pixels in our masterdark, and applied the steps in Sect.~\ref{sec:ipc_method}. The resulting statistics can be found in Fig.~\ref{fig:ipc_kernel}.

\begin{figure}[h]
    \centering
    \includegraphics[width=0.55\linewidth]{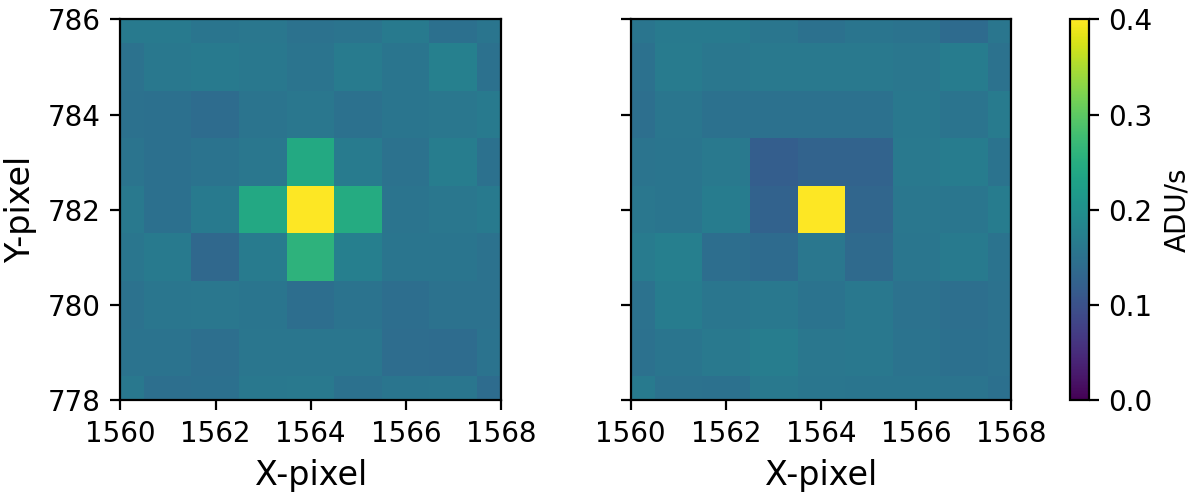}
    \caption{Before (left) and after (right) IPC correction using a deconvolution kernel derived from isolated hot pixels. The characteristic `plus' around the hot pixel is removed.}
    \label{fig:ipc_corr}
\end{figure}

The $\alpha$ values are diagonally and horizontally symmetrical up to the fourth decimal point. However, this is not the case for the direct vertical neighbors. If these values are indeed asymmetrical, they should not be summarized into a single $\alpha_V$. We use the medians presented in the panels for our 3x3 IPC deconvolution kernel, which we apply to correct the data for linear IPC. A before and after around a randomly selected hot pixel can be found in Fig.~\ref{fig:ipc_corr}.

While the `plus'-pattern around the hot pixel is removed, it appears that there is some over-correction above the hot pixel, compared to below the hot pixel. This is interesting, since we retrieved a vertically asymmetrical kernel, with the highest correction value above the hot pixel. The fact that this appears to be over-subtracting the effect, could imply that the kernel should in fact not be asymmetric, and that the IPC is overestimated towards the top of the kernel. We discuss this in more detail in Sect.~\ref{sec:summary}.

\subsection{Non-linearity}
\begin{figure}[h]
    \centering
    \includegraphics[width=0.4\linewidth]{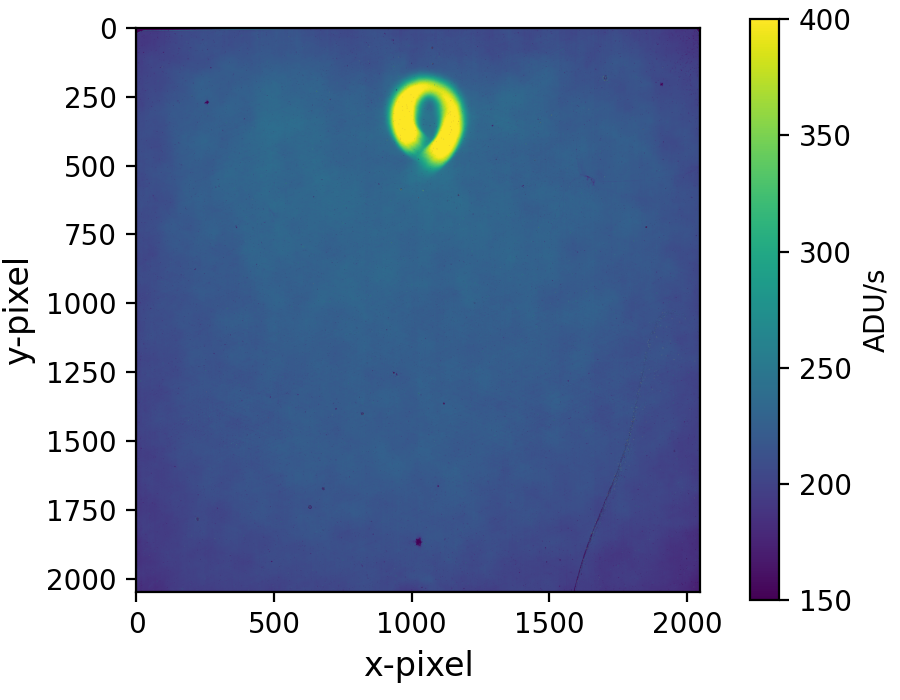}
    \caption{Example of a (corrected) flat taken in the lab. The ring near the top is due to the black body source.}
    \label{fig:flat}
\end{figure}

\begin{figure}[h]
    \centering
    \includegraphics[width=\linewidth]{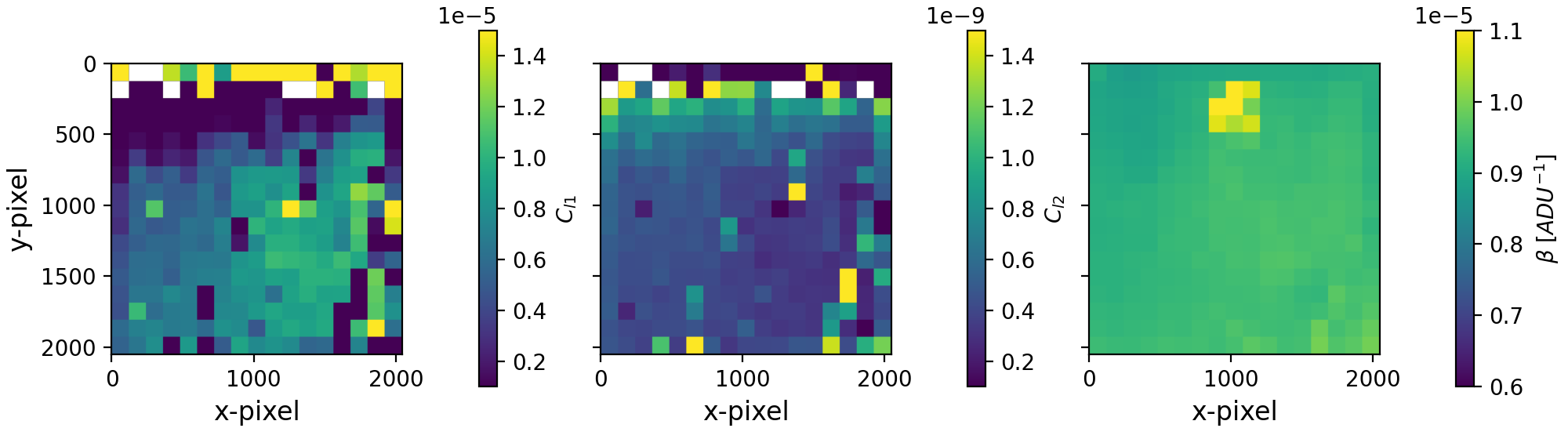}
    \caption{Non-linearity coefficients per super-pixel. The first-order coefficient is shown on the left, the second-order coefficient is shown in the centre. The $\beta$-parameter, which is used in the correlation function, is presented on the right.}
    \label{fig:non-linearity}
\end{figure}

After correcting the darks using the reference pixels and correcting for IPC, we can generate a clean masterdark that can be used for further calibration purposes. The flats were corrected in a similar manner, and the masterdark was subtracted to correct for dark current contributions. An example of a corrected flat is shown in Fig~\ref{fig:flat}. We combine the flats into one average to find the non-linearity correction per super-pixel. Per super pixel, we calculate the results from Eqs.~\ref{eq:ramp_fit}, \ref{eq:lin_fit}, and \ref{eq:nonlin_fit}. The non-linearity coefficients, $C_{l1}$, $C_{l2}$, and $\beta$ are mapped in Fig.~\ref{fig:non-linearity}.

We note a few things regarding the maps. First, the top half of the detector appears strongly affected by the ADC glow. Secondly, there is a clear imprint of the blackbody source in the linearity coefficients. This could imply that migration of charge is already subtly changing the ramp shapes in this area. The coefficients on this part of the detector should probably not be trusted, though we will proceed with the super-pixel results regardless. Finally, larger clusters of bad pixels, such as the diagonal stripe in the bottom right corner, clearly skew the results in a super pixel. Any discussion on the values will be focussed on the more well-behaved and flatly illuminated parts on the bottom half of the detector.

\subsection{The correlation function and charge migration}

\begin{figure}[h]
    \centering
    \includegraphics[width=0.9\linewidth]{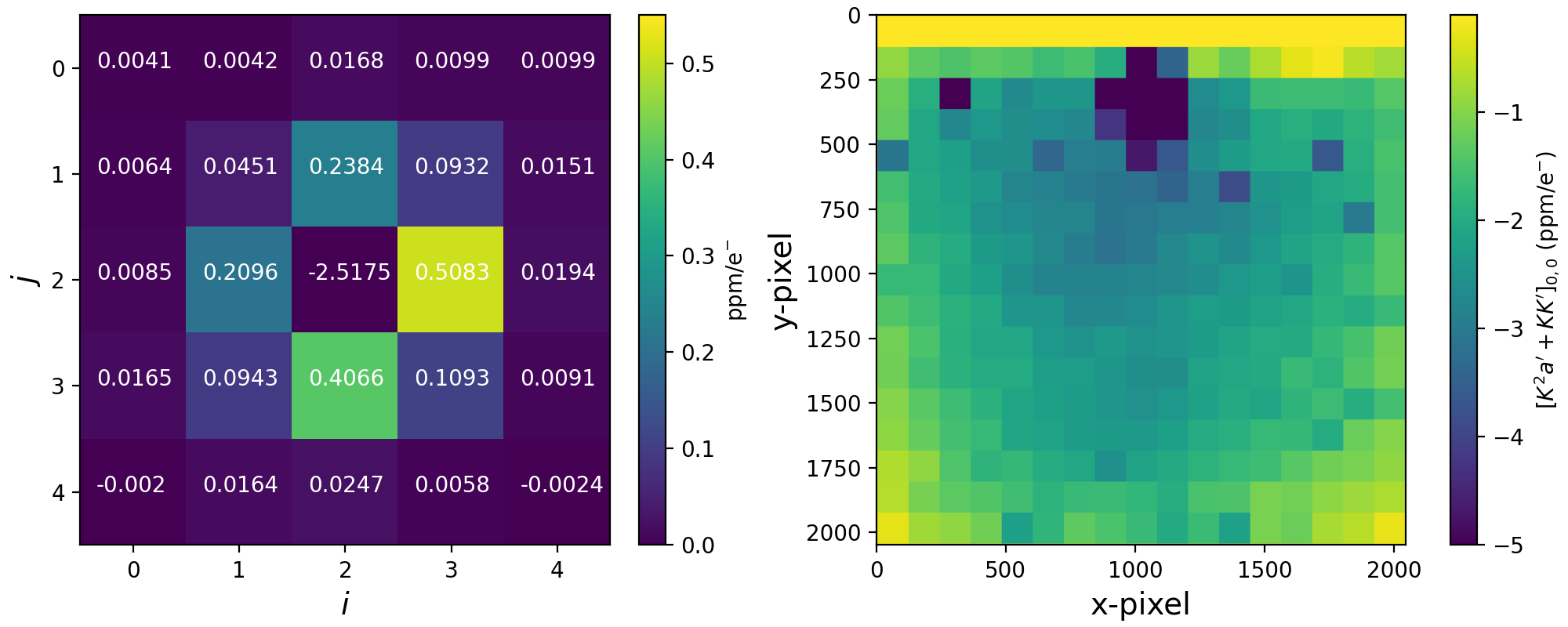}
    \caption{Mean $[K^2a' + KK']_{\Delta i,\Delta j}$ derived from 20 flat pairs and all super-pixels (left), and a map of the mean $[K^2a' + KK']_{0,0}$ (right).}
    \label{fig:bfe_corr}
\end{figure}

We now apply Eqs.~\ref{eq:kfunc} and \ref{eq:corrfunc} to the flats \textit{not} corrected for IPC and non-linearity, as these are folded into Eq.~\ref{eq:kfunc}. We present the results in Fig.~\ref{fig:bfe_corr}, with on the left the mean $[K^2a' + KK']_{\Delta i,\Delta j}$, and on the right the $[K^2a' + KK']_{0,0}$ across the detector. We note that for the mean value of $[K^2a' + KK']_{\Delta i,\Delta j}$, we only used super-pixels around the centre of the detector, due to a significant imprint from non-charge migration related variations in other areas. Similarly to the non-linearity maps, the blackbody source clearly shows up in our map, likely due to it not being flat. This is the same for the edges, which can be seen to be relatively less illuminated than the centre in Fig.~\ref{fig:flat}.

Focussing then on the exact values in Fig.~\ref{fig:flat}, the numbers are roughly similar to the results of other HxRG detectors examined in the works we base ourselves on, albeit on the high side\cite{ref:ChoHiII,ref:20FrGiCh} The central pixel is negatively correlated as expected, since the more signal it accumulates, the more it will spill to its neighbors. The direct neighbors, are the most likely recipients, and therefore are most strongly correlated to the central pixel. The diagonal neighbors still show an excess correlation compared to the correlations at $\pm$2 pixels, indicating they are still relatively strongly affected by charge migration. We do note a significant asymmetry between the bottom right and the top left of the kernel, which may be connected to the asymmetry observed in the linear IPC component.

\section{FUTURE WORK AND OUTLOOK}
\label{sec:summary}

In this work, we have used lab measurements of UTR flats and darks taken by the METIS Imager H2RG and performed a first characterization of the charge migration effect, following essential work developed for the \textit{Roman Space Telescope} detectors and other instruments.\cite{ref:HiChoI,ref:ChoHiII,ref:20FrGiCh} It is becoming an increasingly noticeable problem for many telescopes that does not just affect HxRGs, but also CCDs \cite{ref:14AnAsDo}, and Si:As IBCs\cite{ref:23ArLaRi,ref:24GaArMo}. For METIS, with its HCI and high-resolution spectroscopy modes, this may also prove to be a stringent problem, and we aim to develop at least first-order corrections to limit its effect on the science performance.

It is clear that cross-talk affects HxRGs, as has been well-established by now. It is therefore not unexpected that we observe significant correlations in the flat fields presented here. The characterization performed here is relatively crude. In fact, higher-order terms may be necessary to fully capture the effect\cite{ref:20FrGiCh}, and iterating over the IPC, non-linearity, and charge migration contributions may be necessary to converge to a final solution\cite{ref:ChoHiII}. Regardless, this first kernel estimate provides the first step towards developing a correction for METIS data, and limit the impact on the science performance. 

The detector test data presented here are the first in a series of exposures. Its behavior was not yet tuned, meaning the numbers presented here are not final. Furthermore, METIS contains not one, but five H2RGs. We aim to perform the same analysis on all five detectors using the final voltage settings. The last detector to be included in the instrument, is a GeoSnap. Since its ROIC causes the detector to exhibit much more linear behavior\cite{ref:23LeAtBo,ref:24BoMeTo}, it is not expected to be significantly affected by charge migration, at least not originating from the same physics as the HxRGs. However, it would still be of interest to verify this using similar tests.

Finally, different tests may yield different results. A more direct way to measure migrating charge, is by performing the test under light with a high-contrast pattern. One such tests may be a spot test, where a series of equidistant spots are illuminating the detector.\cite{ref:18PlShSm,ref:24PlShCh} A test under these conditions was shown to yield different results for the impact of charge migration compared to a flat test.\cite{ref:24PlShCh} However, different detectors were used for each method, meaning various factors could explain this difference. The change in magnitude of migration with contrast may be non-linear. Therefore, exploring different test setups with varying illumination scenes on the same detector would be useful to disentangle detector-inherent differences from the real physics affecting charge migration.



\bibliography{report} 

@ARTICLE{ref:ChoHiII,
       author = {{Choi}, Ami and {Hirata}, Christopher M.},
        title = "{Brighter-fatter Effect in Near-infrared Detectors. II. Autocorrelation Analysis of H4RG-10 Flats}",
      journal = {Publications of the Astronomical Society of the Pacific},
         year = 2020,
        month = jan,
       volume = {132},
       number = {1007},
          eid = {014502},
        pages = {014502},
          doi = {10.1088/1538-3873/ab4504},
archivePrefix = {arXiv},
       eprint = {1906.01847},
 primaryClass = {astro-ph.IM},
       adsurl = {https://ui.adsabs.harvard.edu/abs/2020PASP..132a4502C}
}

@ARTICLE{ref:HiChoI,
       author = {{Hirata}, Christopher M. and {Choi}, Ami},
        title = "{Brighter-fatter Effect in Near-infrared Detectors. I. Theory of Flat Autocorrelations}",
      journal = {Publications of the Astronomical Society of the Pacific},
         year = 2020,
        month = jan,
       volume = {132},
       number = {1007},
          eid = {014501},
        pages = {014501},
          doi = {10.1088/1538-3873/ab44f7},
archivePrefix = {arXiv},
       eprint = {1906.01846},
 primaryClass = {astro-ph.IM},
       adsurl = {https://ui.adsabs.harvard.edu/abs/2020PASP..132a4501H}
}

@article{ref:72Ri,
author = {William Hadley Richardson},
journal = {J. Opt. Soc. Am.},
number = {1},
pages = {55--59},
publisher = {Optica Publishing Group},
title = {Bayesian-Based Iterative Method of Image Restoration},
volume = {62},
month = {Jan},
year = {1972},
url = {https://opg.optica.org/abstract.cfm?URI=josa-62-1-55},
doi = {10.1364/JOSA.62.000055},
}

@ARTICLE{ref:74Lu,
       author = {{Lucy}, L.~B.},
        title = "{An iterative technique for the rectification of observed distributions}",
      journal = {The Astronomical Journal},
         year = 1974,
        month = jun,
       volume = {79},
        pages = {745},
          doi = {10.1086/111605},
       adsurl = {https://ui.adsabs.harvard.edu/abs/1974AJ.....79..745L}
}

@ARTICLE{ref:20FrGiCh,
       author = {{Freudenburg}, Jenna K.~C. and {Givans}, Jahmour J. and {Choi}, Ami and {Hirata}, Christopher M. and {Bennett}, Chris and {Cheung}, Stephanie and {Cillis}, Analia and {Cottingham}, Dave and {Hill}, Robert J. and {Mah}, Jon and {Meier}, Lane},
        title = "{Brighter-fatter Effect in Near-infrared Detectors{\textemdash}III. Fourier-domain Treatment of Flat Field Correlations and Application to WFIRST}",
      journal = {Publications of the Astronomical Society of the Pacific},
         year = 2020,
        month = jul,
       volume = {132},
       number = {1013},
          eid = {074504},
        pages = {074504},
          doi = {10.1088/1538-3873/ab9503},
archivePrefix = {arXiv},
       eprint = {2003.05978},
 primaryClass = {astro-ph.IM},
       adsurl = {https://ui.adsabs.harvard.edu/abs/2020PASP..132g4504F}
}

@ARTICLE{ref:23ArLaRi,
       author = {{Argyriou}, Ioannis and {Lage}, Craig and {Rieke}, George H. and {Gasman}, Danny and {Bouwman}, Jeroen and {Morrison}, Jane and {Libralato}, Mattia and {Dicken}, Daniel and {Brandl}, Bernhard R. and {{\'A}lvarez-M{\'a}rquez}, Javier and {Labiano}, Alvaro and {Regan}, Michael and {Ressler}, Michael E.},
        title = "{The brighter-fatter effect in the JWST MIRI Si:As IBC detectors. I. Observations, impact on science, and modeling}",
      journal = {Astronomy \& Astrophysics},
         year = 2023,
        month = dec,
       volume = {680},
          eid = {A96},
        pages = {A96},
          doi = {10.1051/0004-6361/202346490},
archivePrefix = {arXiv},
       eprint = {2303.13517},
 primaryClass = {astro-ph.IM},
       adsurl = {https://ui.adsabs.harvard.edu/abs/2023A&A...680A..96A}
}

@ARTICLE{ref:24GaArMo,
       author = {{Gasman}, Danny and {Argyriou}, Ioannis and {Morrison}, Jane E. and {Law}, David R. and {Glasse}, Alistair and {Gordon}, Karl D. and {Kavanagh}, Patrick J. and {Lage}, Craig and {Patapis}, Polychronis and {Sloan}, Gregory C.},
        title = "{The MIRI/MRS Library. I. Empirically correcting detector charge migration in unresolved sources}",
      journal = {Astronomy \& Astrophysics},
         year = 2024,
        month = aug,
       volume = {688},
          eid = {A226},
        pages = {A226},
          doi = {10.1051/0004-6361/202450241},
archivePrefix = {arXiv},
       eprint = {2406.10835},
 primaryClass = {astro-ph.IM},
       adsurl = {https://ui.adsabs.harvard.edu/abs/2024A&A...688A.226G}
}

@INPROCEEDINGS{ref:24BoMeTo,
       author = {{Bowens}, Rory and {Meyer}, Michael R. and {Tobin}, Taylor L. and {Viges}, Eric and {Hart}, Dennis and {Monnier}, John and {Leisenring}, Jarron and {Ives}, Derek and {van Boekel}, Roy},
        title = "{Characterization of a longwave HgCdTe GeoSnap detector}",
    booktitle = {X-Ray, Optical, and Infrared Detectors for Astronomy XI},
         year = 2024,
       editor = {{Holland}, Andrew D. and {Minoglou}, Kyriaki},
       series = {Society of Photo-Optical Instrumentation Engineers (SPIE) Conference Series},
       volume = {13103},
        month = aug,
          eid = {1310325},
        pages = {1310325},
          doi = {10.1117/12.3018499},
       adsurl = {https://ui.adsabs.harvard.edu/abs/2024SPIE13103E..25B}
}

@ARTICLE{ref:23LeAtBo,
       author = {{Leisenring}, Jarron M. and {Atkinson}, Dani and {Bowens}, Rory and {Douence}, Vincent and {Hoffmann}, William F. and {Meyer}, Michael R. and {Auyeung}, John and {Beletic}, James and {Cabrera}, Mario S. and {Greenbaum}, Alexandra Z. and {Hinz}, Philip and {Ives}, Derek and {Forrest}, William J. and {McMurtry}, Craig W. and {Pipher}, Judith L. and {Viges}, Eric},
        title = "{Evaluating the GeoSnap 13$\mu$m-cutoff HgCdTe detector for mid-IR ground-based astronomy}",
      journal = {Astronomische Nachrichten},
         year = 2023,
        month = oct,
       volume = {344},
          eid = {e20230103},
        pages = {e20230103},
          doi = {10.1002/asna.20230103},
archivePrefix = {arXiv},
       eprint = {2306.05470},
 primaryClass = {astro-ph.IM},
       adsurl = {https://ui.adsabs.harvard.edu/abs/2023AN....34430103L}
}

@ARTICLE{ref:18PlShSm,
       author = {{Plazas}, A.~A. and {Shapiro}, C. and {Smith}, R. and {Huff}, E. and {Rhodes}, J.},
        title = "{Laboratory Measurement of the Brighter-fatter Effect in an H2RG Infrared Detector}",
      journal = {Publications of the Astronomical Society of the Pacific},
         year = 2018,
        month = jun,
       volume = {130},
       number = {988},
        pages = {065004},
          doi = {10.1088/1538-3873/aab820},
archivePrefix = {arXiv},
       eprint = {1712.06642},
 primaryClass = {astro-ph.IM},
       adsurl = {https://ui.adsabs.harvard.edu/abs/2018PASP..130f5004P}
}

@ARTICLE{ref:24PlShCh,
       author = {{Plazas Malag{\'o}n}, Andr{\'e}s A. and {Shapiro}, Charles and {Choi}, Ami and {Hirata}, Chris},
        title = "{Spot-based measurement of the brighter-fatter effect on a Roman Space Telescope H4RG detector and comparison with flat-field data}",
      journal = {Journal of Instrumentation},
         year = 2024,
        month = mar,
       volume = {19},
       number = {3},
          eid = {P03015},
        pages = {P03015},
          doi = {10.1088/1748-0221/19/03/P03015},
archivePrefix = {arXiv},
       eprint = {2310.01920},
 primaryClass = {astro-ph.IM},
       adsurl = {https://ui.adsabs.harvard.edu/abs/2024JInst..19P3015P}
}

@ARTICLE{ref:17PlShSm,
       author = {{Plazas}, A.~A. and {Shapiro}, C. and {Smith}, R. and {Rhodes}, J. and {Huff}, E.},
        title = "{Nonlinearity and pixel shifting effects in HXRG infrared detectors}",
      journal = {Journal of Instrumentation},
         year = 2017,
        month = apr,
       volume = {12},
       number = {4},
        pages = {C04009},
          doi = {10.1088/1748-0221/12/04/C04009},
archivePrefix = {arXiv},
       eprint = {1703.08205},
 primaryClass = {astro-ph.IM},
       adsurl = {https://ui.adsabs.harvard.edu/abs/2017JInst..12C4009P}
}

@ARTICLE{ref:07Ri,
       author = {{Rieke}, G.~H.},
        title = "{Infrared Detector Arrays for Astronomy}",
      journal = {Annual Review of Astronomy and Astrophysics},
         year = 2007,
        month = sep,
       volume = {45},
       number = {1},
        pages = {77-115},
          doi = {10.1146/annurev.astro.44.051905.092436},
       adsurl = {https://ui.adsabs.harvard.edu/abs/2007ARA&A..45...77R}
}

@ARTICLE{ref:14AnAsDo,
       author = {{Antilogus}, P. and {Astier}, P. and {Doherty}, P. and {Guyonnet}, A. and {Regnault}, N.},
        title = "{The brighter-fatter effect and pixel correlations in CCD sensors}",
      journal = {Journal of Instrumentation},
         year = 2014,
        month = mar,
       volume = {9},
       number = {3},
          eid = {C03048},
        pages = {C03048},
          doi = {10.1088/1748-0221/9/03/C03048},
archivePrefix = {arXiv},
       eprint = {1402.0725},
 primaryClass = {astro-ph.IM},
       adsurl = {https://ui.adsabs.harvard.edu/abs/2014JInst...9C3048A}
}

@ARTICLE{ref:15GuAsAn,
       author = {{Guyonnet}, A. and {Astier}, P. and {Antilogus}, P. and {Regnault}, N. and {Doherty}, P.},
        title = "{Evidence for self-interaction of charge distribution in charge-coupled devices}",
      journal = {Astronomy \& Astrophysics},
         year = 2015,
        month = mar,
       volume = {575},
          eid = {A41},
        pages = {A41},
          doi = {10.1051/0004-6361/201424897},
archivePrefix = {arXiv},
       eprint = {1501.01577},
 primaryClass = {astro-ph.IM},
       adsurl = {https://ui.adsabs.harvard.edu/abs/2015A&A...575A..41G}
}

@INPROCEEDINGS{ref:06DoBaSi,
       author = {{Downing}, Mark and {Baade}, Dietrich and {Sinclaire}, Peter and {Deiries}, Sebastian and {Christen}, Fabrice},
        title = "{CCD riddle: a) signal vs time: linear; b) signal vs variance: non-linear}",
    booktitle = {High Energy, Optical, and Infrared Detectors for Astronomy II},
         year = 2006,
       editor = {{Dorn}, David A. and {Holland}, Andrew D.},
       series = {Society of Photo-Optical Instrumentation Engineers (SPIE) Conference Series},
       volume = {6276},
        month = jun,
          eid = {627609},
        pages = {627609},
          doi = {10.1117/12.671457},
       adsurl = {https://ui.adsabs.harvard.edu/abs/2006SPIE.6276E..09D}
}

@INPROCEEDINGS{ref:16DoNiBa,
       author = {{Donlon}, Kevan and {Ninkov}, Zoran and {Baum}, Stefi},
        title = "{Signal dependence of inter-pixel capacitance in hybridized HgCdTe H2RG arrays for use in James Webb space telescope's NIRcam}",
    booktitle = {High Energy, Optical, and Infrared Detectors for Astronomy VII},
         year = 2016,
       editor = {{Holland}, Andrew D. and {Beletic}, James},
       series = {Society of Photo-Optical Instrumentation Engineers (SPIE) Conference Series},
       volume = {9915},
        month = aug,
          eid = {99152I},
        pages = {99152I},
          doi = {10.1117/12.2233200},
       adsurl = {https://ui.adsabs.harvard.edu/abs/2016SPIE.9915E..2ID}
}

@MISC{ref:11HiMc,
       author = {{Hilbert}, B. and {McCullough}, P.},
        title = "{Interpixel Capacitance in the IR Channel: Measurements Made On Orbit}",
 howpublished = {WFC3 Instrument Science Report 2011-10, 18 pages},
         year = 2011,
        month = apr,
        pages = {10},
       adsurl = {https://ui.adsabs.harvard.edu/abs/2011wfc..rept...10H}
}
\bibliographystyle{spiebib} 

\end{document}